\documentclass[referee]{jfm}
\usepackage{color}
\usepackage{amstext}
\usepackage{amssymb}
\usepackage{graphicx}
\usepackage{amsmath}

\newcommand\const{\mathrm{const}}

\newcommand\vV{\boldsymbol{V}}
\newcommand\vI{\boldsymbol{I}}

\newcommand\vf{\boldsymbol{f}}

\newcommand\vl{\boldsymbol{l}}

\newcommand\vx{\boldsymbol{x}}

\newcommand\vg{\boldsymbol{g}}

\begin{document}

{\title[Self-propulsion velocity of $N$-sphere micro-robot] {Self-propulsion velocity of $N$-sphere
micro-robot}}

\author[V. A. Vladimirov]
{V.\ns A.\ns V\ls l\ls a\ls d\ls i\ls m\ls i\ls r\ls o\ls v}

\affiliation{Dept of Mathematics, University of York, Heslington, York, YO10 5DD, UK}

\pubyear{2010} \volume{xx} \pagerange{xx-xx}
\date{Sept 12th 2011}

\setcounter{page}{1}\maketitle \thispagestyle{empty}

\begin{abstract}

The aim of this paper is to derive an analytical expression for the self-propulsion velocity of a
micro-swimmer that consists of $N$ spheres moving along a fixed line. The spheres are linked to each other by
the rods of the prescribed lengths changing periodically. For the derivation we use the asymptotic procedure
containing the two-timing method and a distinguished limit. Our final formula shows that in the main
approximation the self-propulsion velocity is determined by the interactions between all possible triplets of
spheres.

\end{abstract}

\section{Introduction and formulation of problem\label{sect01}}

\subsection{Introduction}
The studies of simple micro-swimmers (or micro-robots) represent a flourishing modern research topic (see
\cite{Purcell, Yeomans, Yeomans1, Lefebvre, Koelher, Gilbert, Golestanian, Golestanian1}), which creates the
fundamental base for modern applications in medicine and other areas. In this paper we generalize the
three-sphere micro-swimmer by \cite{Golestanian} to the general $N$-sphere micro-swimmer. It is possible
since the employed two-timing method and a distinguished limit significantly simplifies the analytical
procedure.

\subsection{Formulation of problem\label{sect01}}
We consider a micro-swimmer consisting of $N$ rigid spheres of radii $R_i^*$, $i=1,2,\dots N$ with their
centers at the points $x_i^*(t^*)$ of the $x^*$-axis ($x_{i+1}^*>x_{i}^*$); ${t}^*$ is time. The spheres are
connected by $N-1$ rods of lengths $l_{i}^{k*}=x_k^*-x_i^*$ where our choice is always $k>i$. The masses of
the spheres and the rods (in the Stokes approximation) are zero. The rods are so thin that their interaction
with a fluid is negligible. The lengths of the rods are prescribed as
\begin{eqnarray}\label{constraint}
&&l_{i}^{k*}=L_{i}^{k*}+\widetilde{\lambda}_{i}^{k*}
\end{eqnarray}
where $L_{i}^{k*}$ are the averaged values and $\widetilde{\lambda}_{i}^{k*}(\tau)$ are the oscillations,
which are prescribed as $2\pi$-periodic functions of $\tau\equiv\omega t^*$  with a constant frequency
$\omega$. Asterisks mark dimensional variables and parameters.

In the Stokes approximation the total force acting on each sphere is zero (their masses are zero), hence the
equation of motion for the $i$-th sphere can be written as
\begin{eqnarray}\label{exact-1}
&&\kappa_i^*\dot x_i^*-\sum_{k\neq i} 3\kappa_i^* R_k^* \dot x_k^*/(2l_{i}^{k*})=-f_i^*
\end{eqnarray}
where $\kappa_i^*\equiv 6\pi\eta R_i^*$, $\eta$ is viscosity, dots above the functions stands for $d/dt^*$.
The l.h.s. of (\ref{exact-1}) represents a viscous friction, while $f_i^*$ is the force exerted by the rods
to the $i$-th sphere. In order to derive (\ref{exact-1}) we have used the fact (see
\cite{Lamb,Landau,Moffatt}) that a sphere of radius $R_k^*$ and position $x_k^*$ moving along the $x^*$-axis
with  velocity $\dot x_k^*$ creates at the center of $i$-th sphere the $x^*$-component of fluid velocity
equal to $-3R_k^*\dot x_k^*/(2l_{i}^{k*})$, where the minus sign corresponds to $x_k>x_i$. The considered
mechanical system is a closed one, hence the total force exerted by the constraints is zero:
\begin{eqnarray}\label{exact-1a}
&&\sum_{i=1}^Nf_i^*=0
\end{eqnarray}
Eqns. (\ref{exact-1}),(\ref{exact-1a}) represent the system of ODEs to be solved in this paper. Notice that
we do not use the summation convention.

The equation (\ref{exact-1}) and its solution $\vx^*(t^*)=(x_1^*, x_2^*,\dots, x_N^* )$, contain three
characteristic lengths: the radius $R$ of spheres, the distance $L$ between the neighbouring spheres, and the
amplitude $\lambda$ of oscillations of rod lengths; at the same time the only explicit characteristic
time-scale $T$ corresponds to the frequency $\omega$:
\begin{eqnarray}
&& R,\quad L,\quad  \lambda, \quad T\equiv 1/\omega
\label{scales-list}
\end{eqnarray}
The dimensionless variables and small parameters are
\begin{eqnarray}
&& \vx^*=L\vx,\quad L_{ik}^*=LL_{i}^k,\quad R_i^*=R R_i,\quad
 \widetilde{\lambda}_{i}^{k*}=\lambda\widetilde{\lambda}_{i}^{k},\quad  t^*=Tt\label{scales}\\
&&f_i^*=-6\pi\eta RLf_i/T,\quad \varepsilon\equiv \lambda/L\ll 1,\quad \delta\equiv 3R/(2L)\ll 1\nonumber
\end{eqnarray}
Then the dimensionless eqns.(\ref{constraint})-(\ref{exact-1a}) take the form
\begin{eqnarray}\label{exact-1-dless}
&&R_i\dot x_i-\delta\sum_{k\neq i} R_{ik} \dot x_k/l_{ik}=f_i,\quad
\quad R_{ik}\equiv R_i R_k\\
&&l_{ik}=L_{ik}+\varepsilon\widetilde{\lambda}_{ik}\label{constraint-less}\\
&&\sum_{i}f_i=\vf\cdot\vI=0,\quad \vI\equiv(1,1,\dots,1)\label{exact-1aa}
\end{eqnarray}
One should note that `dots' above function in (\ref{exact-1}) and (\ref{exact-1-dless}) correspond to the
dimensional and dimensionless time derivatives correspondingly. The first equation (\ref{exact-1-dless}) can
be rewritten in the matrix form
\begin{eqnarray}\label{exact-1-dless-matrix}
&&\mathbb{A}\dot\vx=\vf\quad \text{or}\quad \sum_{k=1}^N A_{ik}\dot x_k=f_i\\
&& \mathbb{A}=A_{ik}=
\begin{cases}
R_i  & \text{ for }\ i=k,\\[-1mm]
-\delta R_{ik}/l_{ik} & \text{ for }\ i\neq k
\end{cases}\label{matrixA}
\end{eqnarray}

\subsection{Notations}

The variables $\vx=(x_1,x_2, ..., x_N)$, $t$, $s$, and $\tau$ serve as  dimensionless coordinates of spheres,
physical time, slow time, and fast time. We use the following definitions and notations:

\noindent
(i) A dimensionless function $f=f(s,\tau)$ belongs to the class $\mathfrak{O}(1)$
if $f={O}(1)$ and all  partial $s$-, and $\tau$-derivatives of $f$ (required for our consideration) are also
${O}(1)$. In this paper all small parameters appear as explicit multipliers, while all functions always
belong to $\mathfrak{O}(1)$-class.

\noindent
(ii) We consider only \emph{periodically oscillating in $\tau$ functions}
\begin{eqnarray}
f\in \mathfrak{H}:\quad f(s, \tau)=f(s,\tau+2\pi)\label{tilde-func-def}
\end{eqnarray}
where the $s$-dependence is not specified. Hence $f\in
\mathfrak{H}\bigcap\mathfrak{O}(1)$.

\noindent
(iii) The subscripts $\tau$ and $s$ denote the related partial derivatives.

\noindent
(iv) For an arbitrary $f\in \mathfrak{H}$ the \emph{averaging operation} is
\begin{eqnarray}
\langle {f}\,\rangle \equiv \frac{1}{2\pi}\int_{\tau_0}^{\tau_0+2\pi}
f(s, \tau)\,d\tau,\qquad\forall\ \tau_0\label{oper-1}
\end{eqnarray}
where  during the integration we keep $s=\const$ and $\langle f\rangle$ does not depend on $\tau_0$.

\noindent
(v)  The class of \emph{tilde-functions} (or purely oscillating functions) is such that
\begin{eqnarray}
\widetilde f:\quad
\widetilde f(s, \tau)=\widetilde f(s,\tau+2\pi),\quad\text{with}\quad
\langle \widetilde f \,\rangle =0.\label{oper-2}
\end{eqnarray}
Tilde-functions represent a special case of $\mathfrak{H}$-functions with zero average.

\noindent
(vi) The class  of \emph{bar-functions} (or mean-functions) is defined as
\begin{eqnarray}
\overline{f}:\quad  \overline{f}_{\tau}\equiv 0,\quad
\overline{f}(s)=\langle\overline f(s)\rangle
 \label{oper-3}
\end{eqnarray}

\section{Asymptotic procedure \label{sect04}}

The use of $\varepsilon$-dependence of $l_{ik}$ (\ref{constraint}) leads  to the presentation of matrix
$\mathbb{A}$ (\ref{matrixA}) as a series for $\varepsilon\to 0$ (we consider $\delta$ as a fixed parameter)
\begin{eqnarray}
&&\mathbb{A}=\overline{\mathbb{C}}+\varepsilon\delta
\widetilde{\mathbb{A}}'_0+\dots, \quad \overline{\mathbb{C}}_0\equiv{\overline{\mathbb{A}}}_0+
\delta \overline{\mathbb{B}}_0
\\
&&\overline{\mathbb{A}}_0\equiv\text{diag}\{R_1,R_2,...,R_N\},\quad\widetilde{\mathbb{A}}'_0\equiv\begin{cases}
0  & \text{for}\ i=k,\\[-1mm]
R_{ik}\widetilde{\lambda}_{ik}/L_{ik}^2 & \text{for}\ i\neq k
\end{cases}\nonumber
\end{eqnarray}
where we do not present the expression for $\overline{\mathbb{B}}_0$ since it will not affect the answer.

The crucial step of our procedure is the introduction of a fast time variable $\tau$ and a slow time variable
$s$. We take $\tau=t$ (which corresponds to the prescribed oscillations of the rods) and $s=\varepsilon^2 t$.
This choice of $s$ can be justified by the same distinguished limit arguments as in
\cite{VladimirovMHD}; here we present this fact without proof, referring only to the most important
fact that it leads to a valid asymptotic procedure. Therefore we use the chain rule
$d/dt=\partial/\partial\tau+\varepsilon^2\partial/\partial s$ and then we  accept (temporarily) that $\tau$
and $s$ represent two independent variables. The two-timing form of eqn. (\ref{exact-1-dless-matrix}) is
\begin{eqnarray}\label{exact-1-dless-matrix-ser}
&&(\overline{\mathbb{C}}_0+\varepsilon\delta
\widetilde{\mathbb{A}}'_0+\dots)(\vx_\tau+\varepsilon^2\vx_s)=\vf
\end{eqnarray}
where unknown functions are taken as the series
\begin{eqnarray}\label{x-f-ser}
&& \vx(\tau,s)=\overline{\vx}_0(s)+\varepsilon\vx_1(\tau,s)+\dots,\quad
\vf(\tau,s)=\vf_0(\tau,s)+\varepsilon\vf_1(\tau,s)+\dots
\end{eqnarray}
The accepted condition $\widetilde{\vx}_0\equiv 0$ reflects the fact that the large distances of
self-swimming are driven by small self-oscillations. Now we consider the successive approximations of
(\ref{exact-1-dless-matrix-ser}),(\ref{x-f-ser}) in $\varepsilon$:

 \noindent (i) \emph{Terms }$O(\varepsilon^0)$  give $\vf_0\equiv 0$.

\noindent (ii) \emph{Terms}  $O(\varepsilon^1)$ give $\mathbb{C}_0\vx_{0\tau}=\vf_1$; the averaged part of
this equation gives $\vf_1\equiv 0$, while the oscillating part yields
\begin{eqnarray}\label{one-tilde}
&&\mathbb{C}_0\widetilde{\vx}_{1\tau}=\widetilde{\vf}_1
\end{eqnarray}

 \noindent (iii) \emph{Terms} $O(\varepsilon^2)$ give the equation $
\overline{\mathbb{C}}_0\widetilde{\vx}_{2\tau}+\delta\widetilde{\mathbb{A}}'_0\widetilde{\vx}_{1\tau}+
\overline{\mathbb{C}}_0\overline{\vx}_{0s}=\vf_2
$; its averaged part is
\begin{eqnarray}\label{two-aver}
&&\overline{\mathbb{C}}_0\overline{\vx}_{0s}+
\delta\langle\widetilde{\mathbb{A}}'_0\widetilde{\vx}_{1\tau}\rangle=\overline{\vf}_2
\end{eqnarray}
The force $\overline{\vf}_2$ can be excluded from (\ref{two-aver}) by (\ref{exact-1aa}):
\begin{eqnarray}\label{two-aver-a}
&&\vI\cdot\overline{\mathbb{C}}_0\overline{\vx}_{0s}+
\delta\vI\cdot\langle\widetilde{\mathbb{A}}'_0\widetilde{\vx}_{1\tau}\rangle=0
\end{eqnarray}
The self-propulsion with averaged velocity $\overline{V}_0$ means that $\vx_{0s}=\overline{V}_0\vI$, hence
\begin{eqnarray}\label{vel-self-prop}
&&\overline{V}_0=
-\delta\frac{\vI\cdot\langle\widetilde{\mathbb{A}}'_0\widetilde{\vx}_{1\tau}\rangle}{\vI\cdot\overline{\mathbb{A}}_0
\vI}
\end{eqnarray}
where the matrix $\overline{\mathbb{C}}_0$ is replaced with $\overline{\mathbb{A}}_0$ since we consider only
the main (linear) term in $\delta$. Expression (\ref{vel-self-prop}) still contains unknown functions
$\widetilde{\vx}_{1\tau}$ which can be determined from (\ref{one-tilde}) with the use of constrains
(\ref{constraint-less}),(\ref{exact-1aa}). Indeed,  the equation (\ref{one-tilde}) (with linear in $\delta$
precision) gives $\widetilde{\vx}_{1\tau}=\mathbb{A}_0^{-1}\widetilde{\vf}_1$ with
$\mathbb{A}_0^{-1}=\text{diag}\{1/R_1,1/R_2,\dots,1/R_N\}$; it means that
$\widetilde{\vx}_{1\tau}=\widetilde{\vg}$ with the components $\widetilde{g}_i\equiv\widetilde{f}_{1i}/R_i$.
One can see that (\ref{constraint-less}) yields
$\widetilde{g}_k-\widetilde{g}_i=\widetilde{\lambda}_{i\tau}^k$, while (\ref{exact-1aa}) leads to $\sum_i
R_i\widetilde{g}_i=0$. Both these restrictions can be written as one $N\times N$ matrix equation
\begin{eqnarray}\label{constraints-tilde}
&& \mathbb{M}\widetilde{\vg}=\widetilde{\vl}_\tau,\ \
 \mathbb{M}\equiv
\left(
  \begin{array}{cccccc}
    -1 & 1 & 0 &\dots &0 &0 \\
    0 & -1 & 1 &\dots &0 &0 \\
    \dots &\dots &\dots &\dots &\dots &\dots\\
    0 & 0 & -0 &\dots &-1 &1 \\
    R_1 &R_2 &R_3 &\dots &R_{N-1} &R_N\\
  \end{array}
\right),\ \
\widetilde{\vl}\equiv
\left(
\begin{array}{c}
    \widetilde{\lambda}_{1}^2  \\
     \widetilde{\lambda}_{2}^3 \\
    \dots \\
    \widetilde{\lambda}_{N-1}^N \\
    0\\
  \end{array}
\right)
\end{eqnarray}
The substitution of $\widetilde{\vx}_{1\tau}=\mathbb{M}^{-1}\widetilde{\vl}_\tau$ into (\ref{vel-self-prop})
gives us self-propulsion velocity in the matrix form
\begin{eqnarray}\label{vel-self-prop-final}
&&\overline{V}_0= -\delta\frac{\vI\cdot\langle\widetilde{\mathbb{A}}'_0
\mathbb{M}^{-1}\widetilde{\vl}_{\tau}\rangle}{\vI\cdot\overline{\mathbb{A}}_0
\vI}
\end{eqnarray}
where the matrix $\mathbb{M}^{-1}$ is
\begin{eqnarray}\label{M-1}
&&(-1)^{N+1}\Delta\mathbb{M}^{-1}\equiv
\left(
  \begin{array}{cccccc}
    \Delta_1-\Delta & \Delta_2-\Delta & \Delta_3-\Delta &\dots &\Delta_{N-1}-\Delta &1 \\
    \Delta_1 & \Delta_2-\Delta & \Delta_3-\Delta &\dots &\Delta_{N-1}-\Delta &1 \\
    \Delta_1 & \Delta_2 &\Delta_3-\Delta &\dots &\Delta_{N-1}-\Delta &1\\
    \dots &\dots &\dots &\dots &\dots &\dots\\
    \Delta_1 & \Delta_2 & \Delta_3 &\dots &\Delta_{N-1}-\Delta &1 \\
    \Delta_1 &\Delta_2 &\Delta_3 &\dots &\Delta_{N-1} &1\\
  \end{array}
\right)\\
&& \Delta_k\equiv\sum_{\alpha=1}^{k}R_\alpha,\ k\geq 1;\quad \Delta\equiv \Delta_N.\nonumber
\end{eqnarray}
Further calculations show that (\ref{vel-self-prop-final}) can be presented as
\begin{eqnarray}\label{vel-self-prop-final+}
&&\overline{V}_0= \frac{\delta}{\Delta^2}\sum_{i<k<l} \overline{G}_{ikl}\\
&&\overline{G}_{ikl}\equiv
R_iR_kR_l\left(\frac{1}{L_{ik}^2}+\frac{1}{L_{kl}^2}-\frac{1}{L_{il}^2}\right)\langle
\widetilde{\lambda}_{ik}\widetilde{\lambda}_{kl\tau}-\widetilde{\lambda}_{ik\tau}\widetilde{\lambda}_{kl}\rangle\nonumber
\end{eqnarray}
where the sum is taken over all possible triplets $(i,k,l):\ 1\leq i<k<l\leq N$. For the three-swimmer this
sum contains the only term, which coincides with one by \cite{Golestanian}. In general it contains
$N!/[(N-3)!3!]$ terms: for the four-swimmer we already have four triplets $(1,2,3), (1,2,4),
(1,3,4),(2,3,4)$, for the five-swimmer -- 10 terms, while for the ten-swimmer the number of triplets grows up
to 120.

The expressions for $\mathbb{M}^{-1}$ (\ref{M-1}) and $\overline{V}_0$ (\ref{vel-self-prop-final+}) have been
obtained by the explicit calculations for $N=3,4,5$ and by the mathematical induction for any $N$.

Formula (\ref{vel-self-prop-final+}) represents the main result of this paper. According to
(\ref{vel-self-prop-final+}) $\vI\cdot\overline{\vx}_s=\overline{V}_0=O(\delta)$; however physical velocity
is $\vI\cdot\overline{\vx}_t=\varepsilon^2 \overline{V}_0$. Hence the order of magnitude of the dimensionless
physical velocity is $O(\varepsilon^2\delta)$.

\section{Discussion}

1. The explicit formula (\ref{vel-self-prop-final+}) allows one to find the optimal strokes, to calculate the
required power, the efficiency of self-swimming, and all related forces (both oscillatory and averaged).
However the large number of terms in (\ref{vel-self-prop-final+}) makes all these problems rather cumbersome,
and places them out of the scope of this short paper.

2. Our approach (based on the two-timing method and distinguished limit) is technically different from all
previous studies in this area. The results for $N$-sphere swimmer show its analytical strength.

3. The expression (\ref{vel-self-prop-final+}) can be predicted without any calculations, on the base of the
result for $N=3$. Indeed, if we are interested in the main term of the order $\varepsilon^2\delta$, then only
the triple interactions can be taken into account, as they have been described by
\cite{Golestanian}. The additional (to triplets) interactions between four spheres will inevitably produce
the next order term $O(\varepsilon^3\delta)$, which we do not consider.

4. There are some interesting discussions about the physical mechanism of self-propulsion in the quoted
literature. However one can also notice that a similar result does exist for self-propulsion in an inviscid
fluid (\cite{Saffman}) and some physical explanation can be achieved if we replace the term `virtual mass of
a dumbbell' by the term `viscous drag coefficient of a dumbbell'. Say, for a three-sphere swimmer this
coefficient decreases when the distance between two neighbouring spheres (a dumbbell) decreases and then the
third sphere is used to `push' or `pull'. If the reverse motion of the third sphere meets the increased drag
coefficient of the dumbbell, then self-propulsion is achieved.

5. The mathematical justification of the presented results by the estimation of the error in the original
equation can be performed similar to \cite{VladimirovX1,VladimirovX2}.

6. One can also derive the higher approximations of self-propulsion velocity, as it has been done by
\cite{VladimirovX1,VladimirovX2}. They can be especially useful for the studies of motion with
$\overline{\vV}_0\equiv 0$ (say, if all correlations involved to (\ref{vel-self-prop-final+}) are zero). In
this case one can show that self-propulsion can be generated by interactions of four and more spheres.

7. In this paper we consider only periodic oscillations of constraints. The studies of non-periodic
oscillations might represent an interesting problem. An attempt in this direction have been made by
\cite{Golestanian1}. In fact, such generalizations have been already considered for many
different oscillating systems.

\begin{acknowledgments}
The author is grateful to Profs. R.Golestanian, A.D.Gilbert, and H.K.Moffatt for useful discussions.
\end{acknowledgments}

\end{document}